# *StudyU*: a platform for designing and conducting innovative digital N-of-1 trials

Stefan Konigorski[1,2,*], Sarah Wernicke[1,2], Tamara Slosarek[1,2], Alexander M. Zenner[1], Nils Strelow[1], Ferenc D. Ruether[1,3], Florian Henschel[1], Manisha Manaswini[1], Fabian Pottbäcker[1], Jonathan A. Edelman[1], Babajide Owoyele[1], Matteo Danieletto[2], Eddye Golden[2], Micol Zweig[2], Girish Nadkarni[2], Erwin Böttinger[1,2,*]

[1]Digital Health Center, Hasso Plattner Institute for Digital Engineering, University of Potsdam, Potsdam, Germany
[2]Hasso Plattner Institute for Digital Health at Mount Sinai, Icahn School of Medicine at Mount Sinai, New York, NY 10029, USA
[3]I. Department of Medicine, University Medical Center Hamburg-Eppendorf, Hamburg, Germany

*Corresponding authors: stefan.konigorski@hpi.de; erwin.boettinger@hpi.de. Digital Health Center, Hasso Plattner Institute for Digital Engineering, Rudolf-Breitscheid-Str. 187, 14482 Potsdam, Germany

# Abstract

N-of-1 trials are the gold standard study design to evaluate individual treatment effects and derive personalized treatment strategies. Digital tools have the potential to initiate a new era of N-of-1 trials in terms of scale and scope, but fully-functional platforms are not yet available. Here, we present the open source *StudyU* platform which includes the *StudyU* designer and *StudyU* app. With the *StudyU* designer, scientists are given a collaborative web application to digitally specify, publish, and conduct N-of-1 trials. The *StudyU* app is a smartphone application with innovative user-centric elements for participants to partake in the published trials and assess the effects of different interventions on their health. Thereby, the *StudyU* platform allows clinicians and researchers worldwide to easily design and conduct digital N-of-1 trials in a safe manner. We envision that *StudyU* can change the landscape of personalized treatments both for patients and healthy individuals, democratize and personalize evidence generation for self-optimization and medicine, and can be integrated in clinical practice.

# Introduction

A widespread aim in current medical research is to derive personalized treatment strategies, which are at the heart of treating every single patient with the best possible therapies. One motivation underlying this aim is that many drugs are only effective in up to 50% of patients (1–3). Hence, treatment guidelines based on population-level randomized controlled trials (RCTs), which derive the best average treatment, may yield non-effective treatment or side effects in up to 50% of the patients. As another characteristic of traditional RCTs, study participants only provide data for population-level analyses and do not profit from participation in the studies. Population-level RCTs are not meant to provide insights for individual participants. The gold standard design for evaluating individual-level treatment effects are prospective longitudinal N-of-1 trials (4), which are multi-crossover randomized controlled trials of sample size one (5,6). That is, the study participant is administered the treatment(s) of interest over time, according to a predefined set-up with treatment length, duration, treatment blocks, and washout phases. In the literature, N-of-1 trials are also described as single-case experimental designs (7).

N-of-1 trials are well-suited for studies when there is large inter-individual heterogeneity in treatment effects, as well as when subpopulation groups or individuals with comorbidities are of interest, who have been excluded from population-level RCTs. A series of N-of-1 trials can be aggregated to provide population-level estimates of treatment effects with similar efficiency compared to RCTs but requiring smaller numbers of participants (8,9). Historically, there have been local implementations of N-of-1 trials in hospitals in the US, Canada, and Australia (10), series of articles on N-of-1 trials have been published in medical and epidemiological journals (11,12), and networks on N-of-1 studies have been formed[1]. The advancements in digital technologies provide the potential to initiate a new era of N-of-1 trials in terms of scale and scope and have opened up new avenues to offer remote healthcare. Still, N-of-1 trials have not been integrated into mainstream clinical research or clinical practice. Some of the underlying reasons might be that despite recently published guidelines (13–15), there are still considerations of the ethical framework when applying N-of-1 trials in clinical care (16) and most importantly, that there is no platform available that allows for an easy and large-scale implementation of digital N-of-1 trials. As of now, conducting a digital N-of-1 trial generally necessitates the development of a new app.

Here, we present *StudyU*. *StudyU* provides an open-source, free, and easy-to-use platform with a study designer for researchers that allows an easy design, customization, and implementation of N-of-1 trials, as well as a study app that allows participants to partake in these trials without having to set-up user accounts. This promises to allow novel interactions between researchers, trials, and study participants.

---

[1] https://www.nof1sced.org/

# Design and Methods

## Related Work

A number of apps for conducting N-of-1 trials have been introduced. An overview of these existing tools that can be used to perform individual-level studies is given in Table 1. The Trialist app (17) has been used in different N-of-1 trials but is currently not available for download and general use. The N1 app (18) is only available for iOS, in the United States, and does not allow customization and further implementation of studies. Several apps exist that provide functionalities for self-tracking and self-quantification but do not allow for an experimental evaluation of interventions (e.g., mPower (19), ParkinsonmPower2 (19)). Of these apps, N1 and mPower are based on the Apple Research Kit. OpenClinica (20) allows creating and conducting studies but focuses on electronic data capture and data management and does neither report results to participants nor allow a collaborative creation of studies. QuantifyMe (21) is a platform that allows users to choose from a limited and pre-specified set of interventions and outcomes to design a study, without further customization possibilities. TummyTrials (22) and SleepCoacher (23) provide possibilities to choose from a set of specified interventions and investigate their effect on sleep and on food triggers in irritable bowel syndrome, respectively. Finally, PACO (24) and movisensXS (25,26) provide tools to design studies but are missing the main component of N-of-1 trials, in that the study app is only gathering data and the results are not analyzed in-app and reported back to the study participant in the app. PACO has the further restriction that it is only available outside of the EU/Switzerland. Also, all of these mentioned platforms except TummyTrials require user accounts, which can create difficulties in terms of data privacy especially if apps are planned to be used in different countries.

Table 1: Overview of existing apps and platforms that are suitable for gathering individual-level data. Some report the results of the conducted studies back to the user (column "Statistical evaluation of results").

| Name | App Availability | Possible Studies/ Diseases | Platforms | Statistical evaluation of results | Customizable | Able to perform N-of-1 trials | Requires User Account | Link to Software |
|---|---|---|---|---|---|---|---|---|
| Trialist | No | Multiple options (for chronic pain only) | iOS, Android, Web | Yes | Limited options | Yes | Yes | N/A |
| mPower | US only | 1 (linked to Parkinson) | iOS | No | No | No | Yes | https://github.com/Sage-Bionetworks/mPower |
| ParkinsonmPower2 | Yes | 1 (linked to Parkinson) | iOS | No | No | No | Yes | https://apps.apple.com/us/app/parkinson-mpower-2/id1375781575 |
| PACO | non-EU/ Switzerland | Flexible creation | iOS, Android, Web | No | Yes | Yes | Yes | https://pacoapp.com/ |
| movisensXS | Yes | Flexible | Android | No | Yes | Yes | Yes | https://www.movisens.com/en/products/movisensxs/ |
| OpenClinica | Yes | 0 | Web | No | Yes | Yes | Yes | https://www.openclinica.com/ |
| N1 | US only | 1 (linked to cognitive health) | iOS | Yes | No | Yes | Yes | https://apps.apple.com/us/app/apple-store/id1308612958 |
| QuantifyMe | Source code only | 4 | Android | Yes | Limited options | Yes | Yes | https://github.com/mitmedialab/AffectiveComputingQuantifyMeAndroid |
| TummyTrials | Source code only | 4 (linked to IBS) | iOS | Yes | Limited options | Yes | No | https://github.com/tractdb/tummytrials |
| SleepCoacher | Yes | Multiple options (linked to sleep) | iOS, Android | Yes | Limited options | Yes | Yes | https://sleepcoacher.cs.brown.edu/#install |
| **StudyU** | **Yes** | **Flexible creation** | **iOS, Android, Web** | **Yes** | **Yes** | **Yes** | **No** | https://github.com/hpi-studyu/studyu |

## Vision

With *StudyU*, our goal is to attract more study participants and researchers to conduct and participate in N-of-1 trials by reducing the set-up process and implementation efforts. We envision that health science and medical researchers or physicians worldwide can use it to collaboratively design and conduct N-of-1 trials. *StudyU* can therefore serve as a platform to contribute to open, transparent, and reproducible medical science by (i) making the study designs of different designed trials directly available to foster reproducibility and well-designed studies, and (ii) making the anonymized data contributed by the study participants of the platform available for analysis to foster the generation of novel medical insights into health intervention effects on the individual- and population-level. We envision to enable democratization and personalization for evidence generation in medicine and personal self-optimization.

## The *StudyU* Platform

The *StudyU* platform consists of three main parts, as illustrated in Figure 1 (see Supplementary Text 1 for more details on the architecture):

  i.   the *StudyU* designer web application for researchers,
  ii.  the *StudyU* app for mobile devices, and
  iii. the backend where the participant data, study definitions, etc. are safely stored.

*Figure 1: Architecture of the StudyU platform. Multiple researchers can collaboratively design and create studies and publish them. Then, study participants can partake in published studies. Study definitions and participant study data are stored in the backend.*

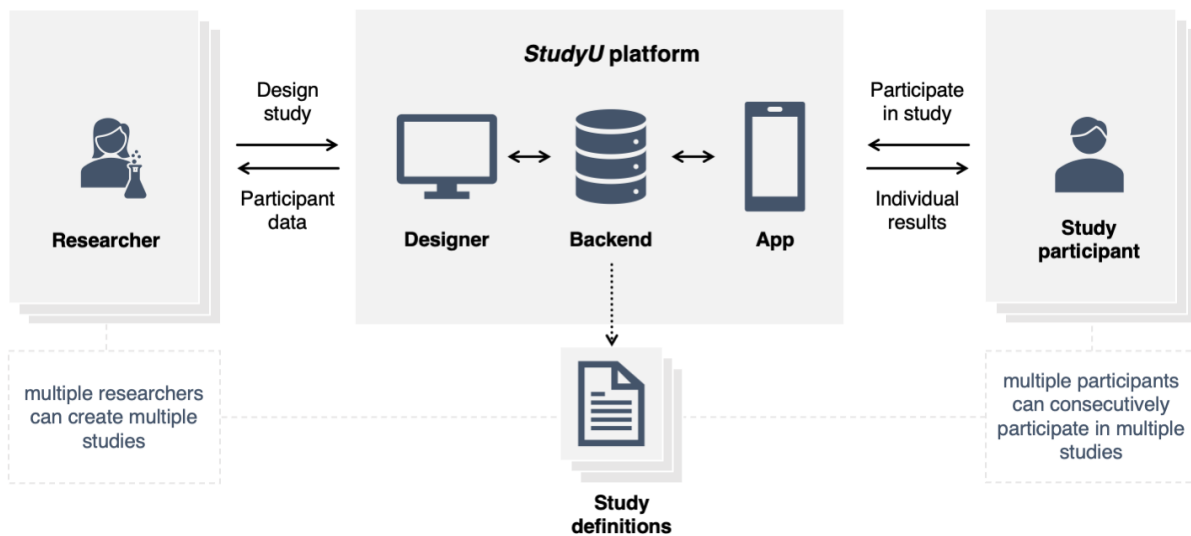

Designing and implementing a study with *the StudyU* designer includes the following steps: specifying the interventions (Supplementary Figure 1), eligibility criteria (Supplementary Figure 2), observations (Supplementary Figure 3) and how they are scheduled, computing the results and displaying them to the participant (see Section 'App'), and consent (Supplementary Figure 4). This user journey is described in more detail in Supplementary Text 2. The designer and the app are currently available in German and English, with Spanish, French and Chinese language apps planned in the near future. In the following sections, the technical setup and the main parts of *StudyU* are described.

## Technical Setup and Use

The *StudyU* frontend applications are written in Flutter[2], an open-source, cross-platform user interface (UI) framework by Google based on the Dart programming language. With this, one single code base can be compiled to performant applications for multiple platforms: mobile, web, and desktop. Parse[3] is used as a backend, which is a platform that incorporates various functionalities such as object storage, user authentication, and push notifications. All components are organized and composed as Docker[4] containers for easy deployment. The source code for the *StudyU* platform is publicly and freely available at https://github.com/hpi-studyu/studyu. For demonstration purposes, the backend is deployed on Back4App[5] and the frontend applications on Google Cloud Run for the designer (at https://studyu-designer-v1.web.app/) and study app (at https://studyu-app-v1.web.app/). StudyU can also be deployed into any HIPAA and GDPR compliant cloud system.

In the current implementation of *StudyU,* two choices can be made regarding how to use the platform, which provides flexibility to the needs of the researcher. First, *StudyU* can either be installed on your own separate server (or cloud) instance, or *StudyU* can be run and accessed on a central server operated by a third party. Second, studies can be designed and published individually, or in collaboration with other researchers from other institutions. For collaborative design, studies can be accessed, edited, and saved by multiple researchers from multiple institutions. The studies can be accessed by multiple researchers at the same time, with the restriction that only one researcher can save data at the same time.

## Study Model

*StudyU* is based on a generic study representation, which is essential to dynamically support multiple different studies. The representation encompasses study metadata and study details. The metadata of a study includes basic information such as the title, a short description, and the researcher's contact, including the institutional review board (IRB) name and protocol number. The study details contain all information that is needed to execute the study: eligibility questions and criteria, interventions, observations, specification of output and report data, schedule, and consent. All objects and relationships are serialized and stored in JavaScript Object Notation (JSON) format. The overall components of this study model are displayed in Figure 2 (see Supplementary Figure 5 and Supplementary Text 3).

---

[2] https://flutter.dev/
[3] https://parseplatform.org/
[4] https://www.docker.com/
[5] https://www.back4app.com/

*Figure 2: Simplified overview of the StudyU study model. The notation is based on the Unified Modeling Language (UML) class diagram notation, which defines properties of single classes in rectangles and associations between multiple classes as connections. The associations shown in this diagram with a filled diamond at one end mean that one class, e.g., 'Study', is composed of another class, in this case, StudyDetails. Numbers shown at associations indicate how many instances of one class take part in this association, e.g., n 'Observation' objects can be associated with one 'StudyDetails' object.*

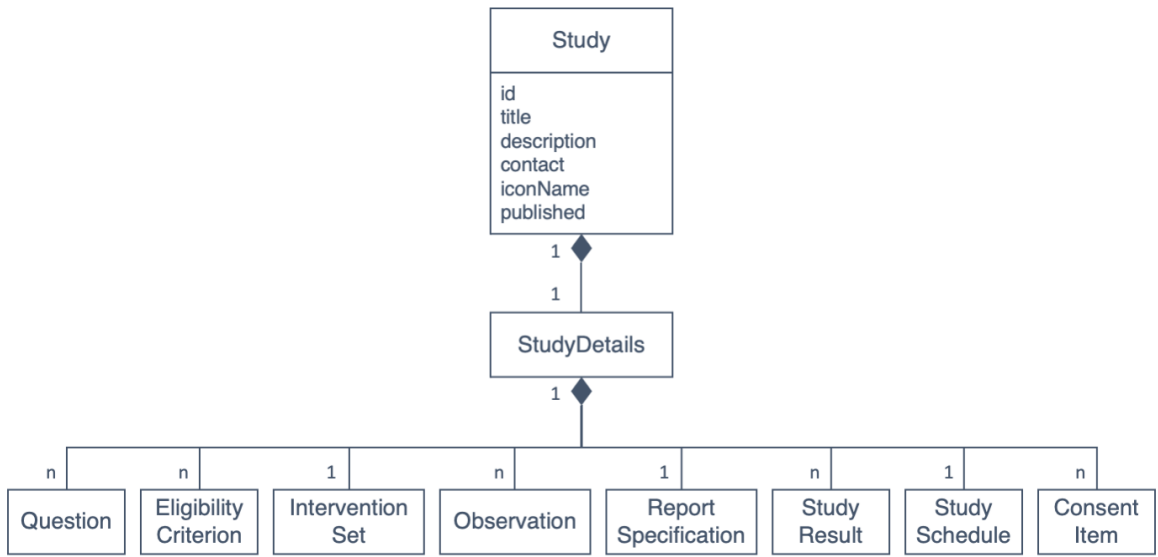

The generic study model allows the design of many different N-of-1 trials in *StudyU*. This is illustrated in two example studies that are implemented in *StudyU*:

1) Investigate and compare the effect of any two of the following daily interventions
    a) willow bark tea
    b) arnica balm
    c) warming pad
   on the intensity of chronic low back pain.

2) Investigate the effect of any two of the following daily interventions
    a) gluten-free diet
    b) low-fibre diet
    c) fructose-free diet
   on the diffuse abdominal pain in irritable bowel syndrome.

## Designer

The *StudyU* designer consists of two main components, the dashboard, and the editor. The rationale behind this concept is to build a user-friendly tool for researchers that provides a logical framework with all the necessary components to plan and conduct a study. Figure 3 shows the dashboard, which displays drafted studies and published studies. Once a study is published, it is available to users in the app and cannot be edited anymore in the Designer. For published studies, researchers can download participant data in CSV format.

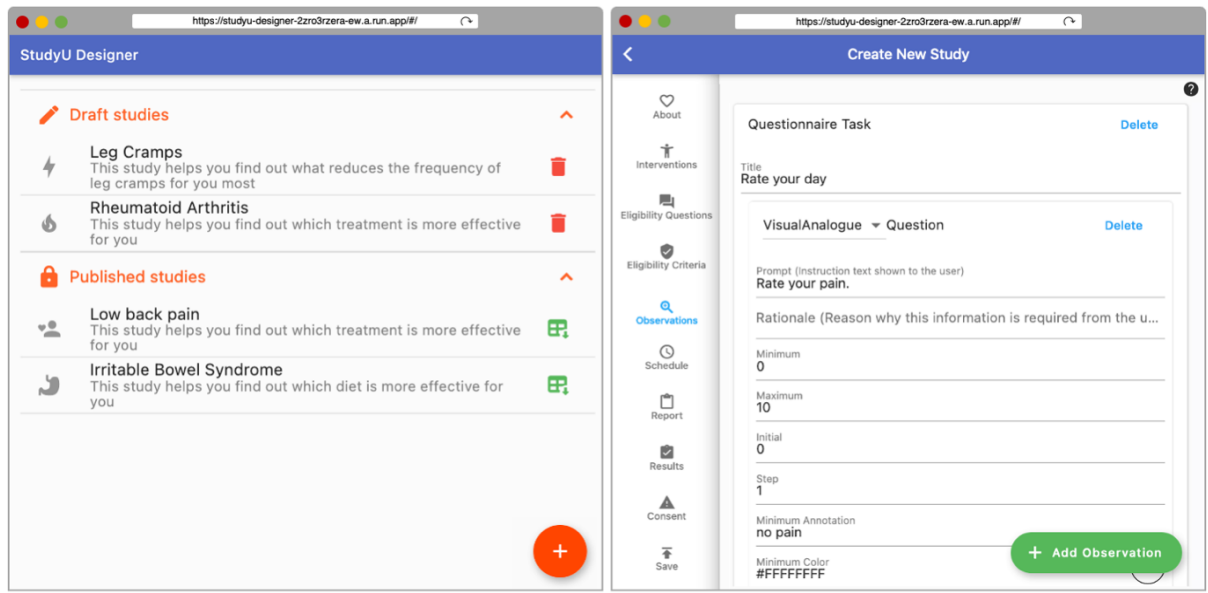

*Figure 3: Dashboard of the StudyU designer with drafted and published studies and an editor screen for observation definition.*

When adding a new study or editing a draft study, the editor leads through all study specifications as defined in the study model, such as interventions, observations, inclusion and exclusion criteria, consent, the format of the downloadable CSV file with study results, and the specification of reports shown to the user in the app. For more editor examples and more details, please see Supplementary Text 4. The sole responsibility for studies lies with the study designers, and in order to ensure the study participants the appropriateness and safety of studies published in *StudyU*, the terms of use of *StudyU* prohibit misuse of the platform and require that researchers have conducted training on good scientific practice. Researchers have to include an IRB protocol number in the study metadata to assure participants of the adequacy of their study.

## App

The app enables users to participate in all the studies that were created and published in the designer. This contains the great advantage that participants do not have to download multiple apps for different studies but can partake in different studies and have an overview of all of them through the *StudyU* app. After the welcome screen (Figure 4a), users can select which of the published studies they want to participate in. Before enrolling for one study, the study metadata is displayed in the study overview screen (Figure 4b). Then, users are led through the onboarding process with a validation of their eligibility, intervention selection, and declaration of consent. Finally, users arrive at the overview of daily tasks (Figure 4c) which contains the study progress bar and the daily tasks (Figure 4d). This is also the default screen users see when opening the app after the initial onboarding.

*Figure 4: First screens of the StudyU app including the welcome screen, study overview screen, and daily screens.*

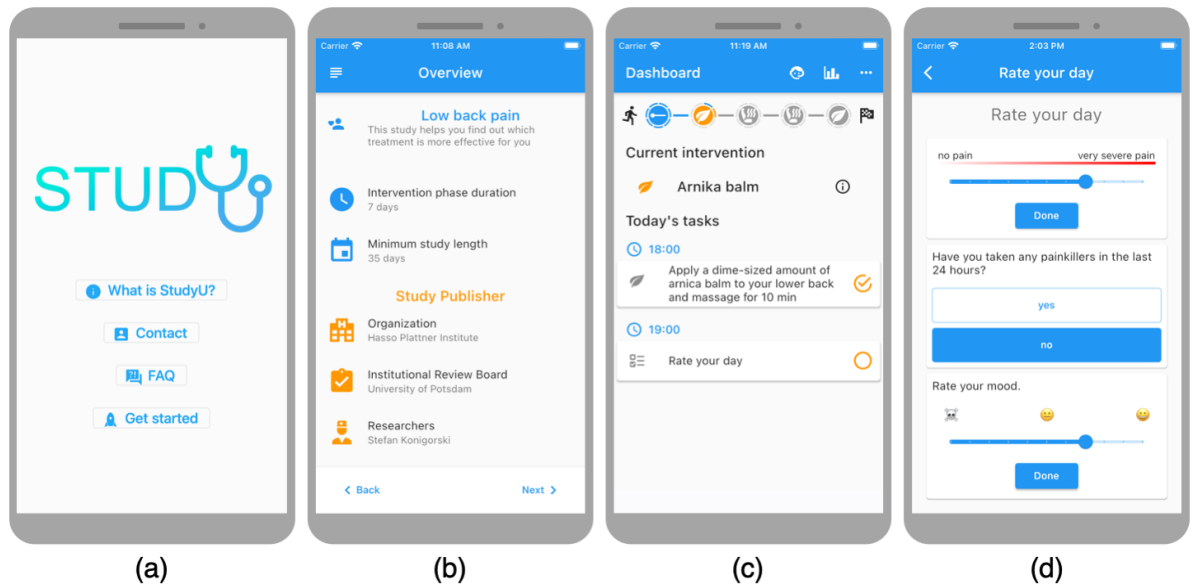

(a)          (b)          (c)          (d)

*Figure 5: Examples of study reports. The power bar in the left top panel indicates whether enough data was collected to observe an effect. Reports are displayed either as a linear regression report or as a report showing data aggregated by day, phase, or intervention.*

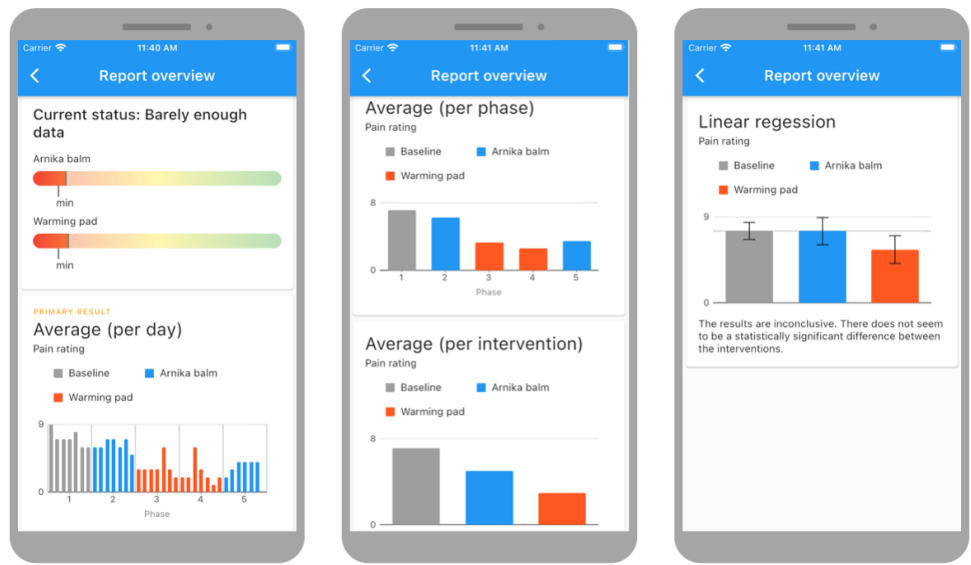

A centerpiece of the *StudyU* app is the result visualization, which is illustrated in Figure 5. In order to ensure that no biases occur after having viewed the results, participants can only view them upon completion of a minimum study length specified by the researcher, which we recommend to specify based on a statistical sample size calculation[6]. Through progress bars, the current status of the participant in the study is visualized to show how many more observations are needed; effects of the interventions (if present) can be expected with the specified statistical power if the participants continue with the intervention at least until they reach the minimum study duration. In the study designer, different report types can be selected: the visualization of a linear regression model that tests if the intervention has an effect on the outcome or a report of individual results are displayed and explained to the participant in bar charts. The definition of report types is implemented in an extensible way. For more details, please see Supplementary Text 5.

---

[6] In the current demonstration of *StudyU* at https://studyu-app-v1.web.app/, the results are available from the first day in order to illustrate how the results look like.

### Data Processing

The studies carried out on the StudyU platform adhere to applicable ethical principles and international regulations, in particular, the General Data Protection Regulation (EU) 2016/679 (27). When the participant opens the app and accepts the terms of service, a new anonymous user account is created with a random ID that is assigned to the participant. In this way, no user profile is needed which could be used to identify the participant. The anonymous account is saved in the backend and on the device. Whenever the app is opened, also the anonymous account gets activated. If the participant completes a daily task, the results are stored inside the UserStudy object and updated on the server. There is the possibility to opt-out of the study, which deletes the unfinished study and the local storage and reference to it. The participant can also choose to be deleted locally and on the server.

The legal basis for processing the study data is the consent provided by the participant via the researcher-defined consent form. Researchers can link and analyze data in different ways in the backend using random user IDs, but cannot link them to specific participants. The setup of *StudyU* does not collect identifiable information, and we also discourage researchers from assessing information in their designed studies that contains person-identifiable data. We anticipate that this set-up without user accounts will satisfy the regulations and data security standards in most countries, allowing a broad use of *StudyU*.

# Discussion

Here, we have presented the *StudyU* platform which allows researchers to undertake an easy design, customization, and implementation of N-of-1 trials, and which allows participants to participate in those without having to set-up user accounts. Through the study designer, researchers can collaboratively design trials. *StudyU* is available open-source, for iOS, Android, and web, is free-to-use and provides anonymized data entry which prevents tracing back the data to the participants. It allows conducting the entire study process digitally, from the study design, participant recruitment, inclusion and exclusion of participants through the study app, automatically analyzing the individual data in the app and reporting the results back to the participant, and saving the data in the secure backend so that researchers can analyze it further and aggregate it across N-of-1 trials. As further innovative concepts, we provide electronic consent and the possibility for the study participants to view their study progress and encourage them to progress further in the study before viewing the results, in order to estimate more precise treatment effects. In the report of the results to the participants in the app, we display a progress bar, which extends the classical statistical-power-based sample calculation. With this, we envision that participants understand the value of long-term participation in the study and stay self-motivated for a longer time so that dropout rates can be decreased. With these features, *StudyU* is currently the only available platform that allows flexibility to N-of-1 trials and conduct them completely digitally. All other existing apps have limitations in the platforms they support, the possibility and customizability to design own trials, the freedom to use the app without having to set-up a user account, and the automated in-app statistical analysis to provide the results back to the participant.

Using the *StudyU* platform, N-of-1 trials can be designed to study the effect of many different health interventions and lifestyle factors on health outcomes in rare diseases and chronic diseases, but also to evaluate the effectiveness of digital health apps. N-of-1 trials can evaluate the effect of health interventions truly in the real-world setting. Especially with the ongoing Covid-19 pandemic where the value of remote and digital medicine is clear, this evaluation and digital integration into the home environment are of high value. While fully digital trials in the home environment can provide challenges for N-of-1 trials due to possible carryover or confounding effects that have to be considered in the automated analysis, the challenges can be addressed through the implementation of more advanced statistical and machine learning methods. In fact, recent years have seen an unprecedented development of deep learning methods for estimating the individual-level effects of health interventions from population-level studies, and to predict individual disease trajectories and individual treatment effects (28–32). These methods are often based on non-testable assumptions, require large datasets, and have limitations in their interpretability of individual treatment effects in complex causal graphs. Combining them with the design advantages of N-of-1 trials can help to derive fully automated analyses of complex real-life trials.

We plan to include several extensions in *StudyU* in the future. First, for the study designer, we plan to add more features encouraging the collaboration on study designs. Setting up a database of interested researchers, clinicians, and institutions can help to search for partners in designing and conducting the studies. As a second feature in *StudyU*, we will include the possibility to include sensor-based measures both as interventions and to assess health outcomes and covariates. Third, we will provide the possibility to design adaptive trials, for example, including elements from micro-randomized trials (33,34) as well as from just-in-time adaptive interventions (35). Fourth, we plan to integrate more complex statistical and machine learning methods in the study app so that complex individual-level treatment effects of potentially time-varying treatments and time-varying confounders can be included in the modeling and results reporting to the individuals. Finally, we are working on the development of user-centric N-of-1 trials designed by the study participants themselves and are excited to integrate these study designs as well as the study results into *StudyU*. Linking them further to electronic health records in the future has the potential to connect N-of-1 trials into clinical care and clinical workflow and can further enhance the integration of medical research and clinical practice.

# Acknowledgements

This work has received funding from the European Union's Horizon 2020 research and innovation program under the grant agreement No. 826117 Smart4Health - Building a citizen-centered EU-EHR exchange for personalized health.

# Author contributions

AZ, NS, MM, FP, FH, FR developed and implemented *StudyU* and *StudyU-Designer* under supervision of SK, TS, SW, EB. SK, SW, TS, EB conceived the project. BO and JE provided critical input regarding design aspects. MD, EG, MZ provide critical input regarding technical and ethical points. SK, SW, TS drafted the manuscript. All authors reviewed and commented on drafts of the manuscript, and approved the final version.

# Competing interests

The authors report no relationships or activities that could appear to have influenced the submitted work.

# Supplementary Material

# Supplementary Text 1: Architecture of *StudyU*

In the following, we give more details on the architecture of *StudyU*.

## Overview

Our system consists of three parts. First, the *StudyU* designer web application for researchers to create studies. Second, the *StudyU* App for mobile devices for participants to conduct the study. Finally, a backend including a database to save the study model, to provide it to the participants and to save the progress of the participants.

The workflow of the system is as follows:

1. The researcher creates and publishes a new study using the designer.
2. The participant selects the study in the App.
3. The participant completes the daily intervention and observation tasks.
4. When the participant finished the study, they can see which intervention improved the outcome for them.
5. The researcher can download the anonymized data of all participants as CSV and use it for further analysis.

## Flutter

Both the App and Designer are built in Flutter. Flutter is a cross-platform mobile application framework. It is being built by Google using their language Dart. The first stable release was published in December 2018. Besides Android and iOS, it supports running as a web application.

Our decision was led by multiple requirements. Flutter allows building an app for both iOS and Android, without developing two separate apps. Flutter has a lot of ready-made components and functionality, developed first party by Google. They provide components in Material and Cupertino design. Flutter also has performance advantages compared to, for example, React Native. Flutter renders all components using Open Graphics Library (OpenGL). This gives it an advantage in performance.

We decided to also use Flutter for the *StudyU* designer, but focused on building a web app. Using Flutter Web has the advantage of reusing parts of the designer, especially the study model classes. Multi-language support and integration with our backend were also reused. We extracted the code into a separate package called core.

## Parse Server Backend

As backend, we used Parse Platform. It provides an easy setup and integration with our app and allows storing studies and user progress. It is an open-source backend focused towards building mobile applications. It was first developed as a service, later bought by Facebook and open-sourced. Parse supports object and file storage (uses MongoDB as a database), user authentication, push notifications and has a dashboard. We used the Parse Flutter Software Development Kit (SDK), which is an almost complete implementation of the Parse API.

We built a Parse setup with docker-compose, which can be run on every machine with Docker installed. The Parse backend stores all data produced by the applications. First when the *StudyU* App is started, an anonymous user is created and used to associate study participation to that user. The user is stored on the device and does not carry any information of the participant. We also store the studies, created by the designer, split into tables *Study* (basic information) and *StudyDetails* (complete study model). Each time a user starts a new study, a new *UserStudy* object is created. This object holds all information that is needed to run the study for that user. It saves study information such as name and description, but also properties set by the user, such as the selected intervention. We duplicate this data to ensure that changes to the study will not interfere with already running studies. Every time a user completes a task, its results are stored on the server. This allows the researcher to always check how the running studies are doing and how many participants the study has. The researcher can also download and analyze anonymized data of participants. Parse supports sending push notifications to the phones. This could be used to – anonymously – encourage users to finish a study or make them aware of certain events without them needing to open the app.

## *StudyU* App

We built the *StudyU* App to run primarily on iOS and Android, but it can also run on the web. The app has multiple instances where it interacts with the Parse server or the internal store. First when the users open the app and accept the terms of service, a new anonymous user account is created. It is saved in the backend and on the device. Each time from then on when the app is opened, we have an active user account until the user deletes the app or deletes its data. After the terms are accepted, the app fetches all published studies from the backend. When the user then selected a study and made their intervention choices, the app creates a new *UserStudy* instance and copies all data needed to run the study from the *Study* and *StudyDetails* instance. Additionally, a reference to the newly created *UserStudy* object is saved inside the local storage. Each time the app is started, we check if a *UserStudy* is present in the local storage and fetch the corresponding object from the backend with it.

Each time the user completes a daily task, the result is saved inside the *UserStudy* object and updated on the server. The app also has the option to opt-out of the study, which deletes the unfinished study and the local storage reference to it and sends the user back to study selection. The user can also press on delete, which deletes the user locally and on the server, but does not delete the current study. This will be changed in the future, for consistency with the opt-out method.

## *StudyU* Designer

The *StudyU* Designer was built using Flutter, with a focus on a web version of the application. The *StudyU* Designer creates the studies and saves them on Parse. Studies have a published attribute. An unpublished study is not seen in the app and can still be edited. Once published the study is seen in the app, but cannot be changed in the *StudyU* Designer anymore. The *StudyU* Designer also supports Android and iOS, which will be added in future versions of *StudyU*.

## Core Package

Since both the App and the Designer use Flutter and Dart, we split out common functionality into a separate package called core. The package is imported both by the App and Designer and holds all the model classes and some other functionality used by both apps. It is a great starting point to build an additional application using the same study model, as all model classes are contained in the package. Furthermore, we could extract more shared code, even User Interface (UI) components, which are needed in both.

# Supplementary Text 2: User Journey

In this section, we provide an overview on the main concepts of the *StudyU* platform. For each step, the view of the researcher who designs the study and the user who participates in the study are shown.

In the interventions section (Supplementary Figure 1), the researcher is able to specify treatments or therapies that will be applied in the study over time at specified time intervals, by using text and icons. These interventions are safely stored in the backend, and then the study participant can choose from them for the study. Details are displayed under the information icon and also the involved daily tasks are shown here. It is not possible to select more than two interventions per study.

In N-of-1 trials and clinical trials in general, the definition of inclusion and exclusion criteria for participation in the study is of particular importance. This can be done in the designer through the two sections 'Eligibility Question' and 'Eligibility Criteria' (Supplementary Figure 2). The eligibility criteria can be flexibly set up based on the answers to one or multiple questions. For the App, an explanation can be provided to the participant explaining the exclusion criteria.

Supplementary Figure 3 shows how observations can be defined in *StudyU*, i.e., all primary and secondary outcomes, as well as further covariates of interest such as lifestyle or environmental factors. The current version supports two task types: tasks that can be checked upon completion and questionnaire tasks. The modularity of the app also allows the extension of further task types. For questionnaire tasks, different question types such as multiple-choice questions or logical statements can be selected as well as rating scale questions. Annotated gradient colors, appealing texts and emojis help to enhance the user experience. After completing this section, the researcher can specify the schedule. The study length can be defined under 'StudySchedule' where the desired number of cycles, duration of phases in days and a baseline phase can be defined. Further, the sequence of interventions can be specified as alternating, counterbalanced or randomized. In the *StudyU* app, the participant gets an overview of the final study schedule with a detailed study description and information about the responsible researchers. The start date of the trial is shown for each study phase and the last entry in the list marks the final trial session after which the study results can be viewed.

As we value a good understanding of study consent by the study participant of great importance, we aimed to define and convey the consent display in a comprehensible and appealing way using modular, topic-specific boxes (see Supplementary Figure 4). The rationale behind the design of the screens is to make the consent process stripped-down and as attractive as possible. This is realized through icons and animations that can be added to the boxes. Study participants can trace which modules they already read through the changing colours of already opened boxes. At the end of the consent process the participant agrees or refuses to take part in the study. The content can also be saved on the participant's device.

# Supplementary Text 3: Study Model

In the following, we give more details on the study model, which is shown in Supplementary Figure 5.

## Overview

One desired goal of the project was development of a platform that can handle multiple different studies. This raises the question of how to distribute the studies to the user. The studies should be executed using a mobile app available in the respective app store of the device. However, if an update would be required for every new study, the deployment of a study would take a very long time due to review processes in the app stores. Hence, the app should be capable of executing new studies without the need of updates. To solve this problem, we treat study designs as files that are stored on a server. When a researcher designs a new study, the design is made available to all users of the App. A user can then select the study to start participating in it and gets also the study contact information displayed. The design is downloaded by the app and executed according to the ideas of the researchers. Once the user finishes the study, they can select another study to participate in.

To allow for the representation of studies as files, we defined a data model for N-of-1 trials which covers all common elements of N-of-1 trials. This model consists of a set of objects and their relationships and is serialized using JavaScript Object Notation (JSON). It is split into two parts, metadata and details. The metadata of a study contains its title, a description, the research group and other aspects that are shown when the user looks at the list of available studies. The details of a study contain the study description and all information that is needed to execute it. These details are further separated into multiple aspects which map to the common elements of N-of-1 trials.

## Elements and Aspects

An N-of-1 trial consists of interventions and observations. In the model, the interventions are stored as an *InterventionSet* which stores a list of *Intervention* objects. The observations are represented as a list of *Observation* objects. The desired schedule of the interventions in the study is stored as a *StudySchedule* object. This contains the desired number of cycles and the duration of a phase in days. Further, the inclusion of a baseline phase can be configured. From these values, the total duration of the study can be calculated. The sequence of interventions can be specified as alternating, counterbalanced, or randomized.

The exclusion criteria are split into two elements, an eligibility questions and a set of eligibility criteria. The eligibility questions consist of a list of *Question* objects. The eligibility criteria are Boolean expressions that reference the answers to questions. If any criterion evaluates to false, the user is excluded from the study and provided the reason specified in the model. If all criteria evaluate to true, the user is eligible to participate in the study. To provide facilities for informed consent, the consent process is broken down into multiple items. These items might include risks of the study, privacy considerations, and funding information. Each of these items is represented as a *ConsentItem*.

To provide understandable results to the user and facilitate further analysis by researchers, the outcomes observed during the study are processed in two ways. First, a visual and easily comprehensible report is generated for the user based on a *ReportSpecification*. Second, the aggregated and anonymized results can be exported by researchers for the analysis in other tools as specified by a set of *StudyResult* objects. We will further examine two aspects in detail, the *Question* objects and the *ReportSpecification*.

## Eligibility Questions

As stated above, an eligibility questions item consists of a list of *Question* objects. When eligibility questions are presented to a user, the questions are presented in order. There are four types of questions to cover different types of data to be collected. First, a *BooleanQuestion* covers yes or no questions such as "Have you had back pain in the last 12 weeks?". Second, a *ChoiceQuestion* provides single-choice or multiple-choice questions which can be used to inquire about gender or present allergies. Finally, a *VisualAnalogueQuestion* and *AnnotatedScaleQuestion* inquire numerical values in a range via a slider. The difference between the last two types is the scale that is displayed next to the slider.

The answer to a question is stored in an *Answer* object. Based on a set of answers for eligibility questions, *Expression* objects are used to represent Boolean expressions for the eligibility questions. As stated previously, these expressions are used to determine the eligibility of the user. Further, they can be used to make some questions optional. One example for this is the exclusion of the question "Are you pregnant?" if the user already specified that they are male. If a question is excluded because of such a condition, a default value is used instead.

## Reports

As stated previously, interventions and observations consist of tasks. Each type of *Task* can produce results which are stored every time the task is performed. The type of results includes whether the user completed a simple task or answers to a daily questionnaire. These results are associated with the intervention that was active at the time for the purpose of analysis.

For the visual presentation of results to the user, a *ReportSpecification* is used as a blueprint. A report consists of at least one primary *ReportSection* and optional multiple secondary sections. There can be different types of sections that reference the results generated over the course of the study. One simple example for a section is the *AverageSection*. This should present a single observed outcome averaged over a configurable time frame (either per day, per phase, or per intervention).

As there can be many different kinds of tasks, the format of results of these tasks can vary widely. To provide a configurable method to reference observed outcomes of different tasks, *DataReference* objects are used. These data reference objects specify a type of data they refer to (e. g. numerical values). A reference consists of a task identifier and a property identifier. A task might expose only one property (the user completed the task) or multiple properties (the answers to a questionnaire). When resolving a data reference to generate a report, the type of the property is checked and a list of values with accompanying timestamps is exported. The values can then be used by the report section to present the data to the user.

# Supplementary Text 4: *StudyU* Designer

In this section we describe the *StudyU* Designer, the app that lets researchers design and publish studies. We motivate why this app is needed and look at its components, including a more detailed description of certain features it provides along with some of the highlights of its implementation.

## Motivation

As introduced in the previous section, the study model we designed generalizes N-of-1 trials and offers the researchers a lot of flexibility to create their own studies just the way they envision them. However, the following question arises: How can the researchers incorporate their ideas on how the study they want to conduct should look like? And how can they deploy the final study on our platform? One solution would be to simply allow the researchers to upload a study according to the schema of our Study Model and in the JSON file format that our platform uses to store the studies. With this a validation check of the JSON against our study model would suffice. This approach shifts a large amount of work to the researchers, and would require more technical expertise. Instead, we wanted to achieve the goal of building an easy-to-use app for the researchers, that provides a logical order of defining the different parts of a study as well as the required building blocks that can be configured with relative ease. Our hypothesis is that by reducing the friction of the setup process, ultimately more researchers will be willing to conduct N-of-1 trials using our platform. This in turn makes the *StudyU* App more attractive to its users, as it provides more studies to participate in.

## Components

The Designer can be divided into two main components, the dashboard and the editor.

### Dashboard

The dashboard shows the researcher two lists of studies: draft studies, which are studies that are still being worked on, and published studies, which are studies that users can participate in. Draft studies can still be deleted by the researcher. For published studies, the results can be downloaded.

### Editor

The editor allows the researchers to specify the different aspects of their study according to our previously introduced study model. The editor is split further into sections that allow the user to focus on each aspect separately and to have a better overview. Also, we have arranged these sections in an order that represents a logical way of approaching the setup of a study. However, the order is not mandatory and the researchers can switch between the sections as they please. In the following we will introduce the sections briefly in the same order as they are presented in the Designer.

a. <u>Meta Data</u>: This section allows the researcher to edit the study title, description and select a descriptive icon to represent the study.

b. <u>Interventions</u>: Here the researcher can define the different interventions that are offered to the user, including the tasks each intervention comprises and how these tasks are to be scheduled. We put this section second, just after the Meta Data section, as the interventions are the main characteristic of a study and we expect them to impact the other setup decisions.
c. <u>Eligibility Questions</u>: In the eligibility quesionts, the questions the user needs to answer before participating in the study are added. The researcher can specify Boolean questions as well as single and multiple-choice questions.
d. <u>Eligibility Criteria</u>: Using the questions defined in the section before, the researcher can now define the eligibility criteria by specifying what answers lead to an exclusion, together with a reasoning for the users on why they are excluded. This section can only be edited once the eligibility questionnaire above contains at least one question.
e. <u>Observations</u>: In this section the researcher defines the observations the user has to report on and how they are to be scheduled. In the current version the user can be asked to answer different questions. Additionally, to the Boolean and choice questions already mentioned, the researcher can also set up slider questions.
f. <u>Schedule</u>: The schedule section is used to edit the study's schedule information, including the number of cycles, phase duration, whether or not a baseline phase should be included and the sequence of the interventions. Different interventions have different impacts, e.g., a longer washout period or a longer time until an effect can be observed.
g. <u>Report</u>: Here the researchers define the setup of the report the user receives. They can add and configure the different report sections.
h. <u>Results:</u> This section allows specifying which observation results the researchers want to include in their data export.
i. <u>Consent</u>: In the consent section, the researcher can set up the study-specific consent form in a modular way. Placing the consent section last allows the researchers to take all their previous settings into account. For instance, the researcher might be using all, a selected few or none of the observation results, each option impacting the content of the consent.
j. <u>Publish</u>: The final section offers the researcher the options to save the study as a draft or to publish it. If saved as draft, the researcher can continue editing the study at a later point in time before eventually publishing it. When the researcher wants to publish the study, it is first validated, to check whether all necessary aspects are specified. Once published it is available to the users of the *StudyU* App.

# Features

In the following, we highlight some of the features we included to make the Designer easier to use.

### Inline Editing & Icon Picker

As can be seen in the actively edited title field below, the researcher just needs to click into a field to edit its value, instead of having to open an edit model or having to navigate to an edit page.

[Study title/description form image]

We offer the researcher to select icons, for the study itself, its interventions and the consent items as visual clues for the user.

## Type Changing

The study model offers multiple types for certain elements of a study. As an example, a question can be a Boolean, choice or slider question. This poses the question how the researcher can easily choose the desired type while setting up the study. One option would be to offer multiple add buttons, one for each type. However, that would make the interface more complicated. Instead, we only offer one add button and initially add the most basic type of an element, in case of the question element that is a Boolean question. Then the researcher can easily switch the type with a drop-down button. The type can also be switched after the element is already edited. When changing the type, we keep the values for shared attributes between the old and new type of an element, such as the prompt field shown below. This saves the inconvenience of having to type things again.

[Before/After comparison of question type switching]

Before                 After

## Nested Expressions

When defining the eligibility criteria, the researcher defines via expressions what eligibility answers are required for the user to be eligible. By nesting a value expression inside a not expression, a negation can be achieved. The nesting is also supported visually in the designer as shown below. In the shown example the criterion defines that a participant must answer the "Are you pregnant?" question with "no" to be eligible.

## Slider Questions

One of the best examples of the amount of flexibility our study model and therefore the *StudyU* Designer offers the researcher is the configuration of slider questions. The researcher can define the minimum, maximum and initial value, as well as the step size of the slider. In case of a visual analogue question, annotations for the minimum and maximum value can be added and even the color gradient is customizable. The figure below shows the section of the editor for these values in the designer and how the question is displayed to the user in the app. It illustrates that the color gradient can be configured to underline the intention of the question, in this case choosing between the most positive and most negative option by sliding along a white to red gradient.

StudyU Designer

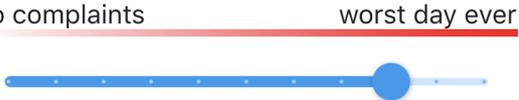
StudyU App

# Supplementary Text 5: *StudyU* app

## App Onboarding

This part of the app is the only part accessible directly after installing the app. Its purpose is the introduction of the app's objective and of the app's terms and conditions.

### Welcome Screen

The welcome screen is the first screen the users reach after opening the app. It is kept clean to not overwhelm the users. The only elements are our app logo, a button to go to the next screen, the terms and conditions, and a button to reach the about screen if users want to get more information about the app and N-of-1 trials before starting a study.

### About Screen

The about screen is intended to explain the concept of the app to users in a way that makes them want to participate in a study. Therefore, in our version the users are introduced to the idea of N-of-1 trials by means of an everyday example. In order not to overload the screen with text, single scrollable pages were created, which are rounded off by individual icons for clarification.

### Terms and Conditions Screen

This screen displays the terms and conditions the users need to accept in order to use the app. The contents of this screen can be saved to the user's device with a button at the bottom of the list. The file gets created with the currently selected language and saved to Downloads on Android and to a separate folder in Data on iOS. The user can finish the onboarding and reach the study selection if all terms are accepted by clicking the checkboxes.

## Study Onboarding

After the user was introduced to the app and the terms of service, the next step is starting a study. This process begins with selecting a study, checking for eligibility and configuring it to the user's interests. The users are also taken to the beginning of this part of the app, after aborting a study selection, after opting out of a study or after finishing a study and starting a new one. During this onboarding, the users can navigate back and forth through the steps with buttons at the bottom of the screen and progress bars how the progress of the onboarding from the eligibility checks onward. This is designed to make the process transparent and keep navigation uniform.

### Study Selection Screen

The study selection screen is the start of every study. It displays a list of available studies with their name, a representative icon chosen by the researcher and a short description of the study. After selecting a study, the user can start the study specific onboarding with the study overview screen.

## Study Overview Screen

This screen offers an overview over the study the user chose. The main points are the length of the study phases, the (minimum) duration of the study, a description of the study and information about the researchers responsible for the study. This screen has further potential to be extended with additional information the researchers want to provide to the users about their study. There is no interactive element on this screen and the users can continue to the eligibility check questionnaire.

## Eligibility Check Screen

The eligibility check is used to prevent users from taking part in a study they are not suitable for. The questions for the check get added to a list one at a time, after the users answered the previous question. This allows the users to focus on one question and also to change an answer to an earlier question in case they made a mistake. When the users correct an answer, the eligibility questions get reset to the corresponding question to keep that step-by-step flow. The questions can be interdependent so that certain questions don't get asked if they are not necessary based on the answer to a previous question. The supported questions are simple yes or no questions or choice questions with either multiple possible selections or a single one. If the users give an answer that disqualifies them from taking part in the study, a popup informs the users about that fact with the given reason and proposes correcting an eventual mistake or going back to the study selection. This is shown in the figure below. If nothing prevents them from taking part in the study, after the last question has been answered the popup will allow the users to continue to the intervention selection.

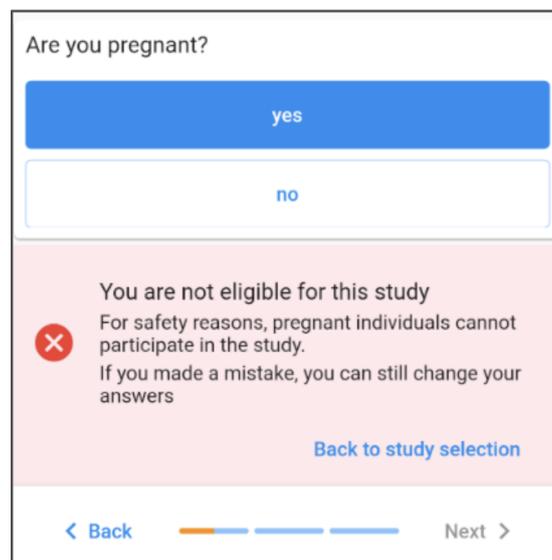

## Intervention Selection Screen

This screen allows the users to customize the study according to their interests by choosing two interventions that will be compared during the study. The interventions can be selected from a list defined by the researcher and are presented in a short way with their icon, name and the corresponding daily tasks ordered by time. The users can get further information by tapping an information icon next to the name of the intervention that contains the description of the intervention. After two interventions have been selected (no more than two interventions can be selected), the users can progress to the journey overview screen.

### Journey Overview Screen

The journey overview screen displays the final study schedule of the interventions to the users based on the previously selected interventions and the sequence that was set by the researcher. The starting date is displayed next to each study phase and the last entry of the list is the final date when results will be available. This screen is followed by the consent screen.

### Consent Screen

The consent screen is modular in the form of topic-specific boxes. The number and content of the boxes (text, images, animated video) can be defined by the researcher in the *StudyU* Designer. The rationale behind the design of the screen is to make the process of education and consent screen as minimal and at the same time as attractive as possible. Therefore, icons or animations can be added to the boxes and users can recognize which box has already been opened by its color that changes after clicking on it. At the end of the process, users have the opportunity to refuse to participate in a study or to give their consent. The implementation requires users to have read the complete content and agree to it in order to give consent and proceed further. The consent can also be saved on the user's device with a button in the upper right corner of the screen, similar to the terms of service. By choosing to continue, the user finishes the onboarding process and gets redirected to the dashboard to start the study.

## Study Participation

The screens of this part of the app are the main part that is accessible during a running study. They allow the user to do the study tasks, look back into the results of past studies and configure some basic app settings. After a study has been started, the app will also send reminders when tasks are supposed to be done as shown below.

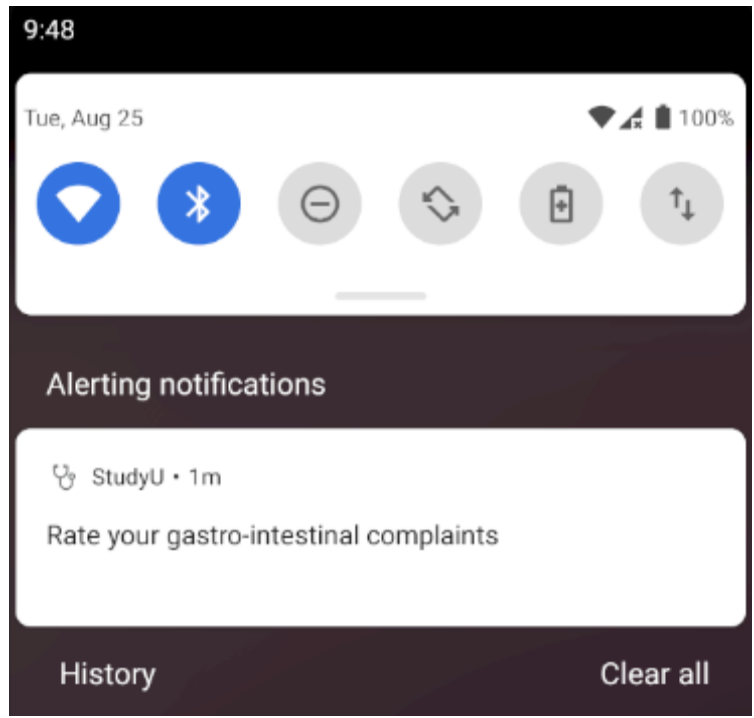

### Dashboard Screen

The dashboard screen is the main screen for the user to access the app's functionality. The focus of this screen is the list of tasks the user has to complete to progress with the currently selected study. The tasks are represented with small cards to allow grouping by the time the tasks shall be conducted. The cards only consist of the title of the tasks and a completion status checkbox to allow for a compact overview even if many tasks are present. At the top of the screen is a timeline with the different intervention icons surrounded by indicators showing how many days have been completed during each phase. Below that timeline is the name of the current intervention and next to it is an information icon that contains the interventions description. These elements display the current study state in a concise way. At the top of the screen, the users can reach the contact screen, the report history and the settings.

### Task Screen

The task screens are used by the users to complete study tasks. Currently, the two supported task types are checkmark tasks and questionnaire tasks. This aspect of the app is modular and can easily be extended with more task types. The checkmark task screen contains instructions on how to complete the task and a button to confirm completion. The eligibility questions screen contains a list of questions similar to the eligibility check. In addition to the multiple-choice questions and Boolean questions, this eligibility questions also support slider questions that allow the user to rate an aspect on a scale according to their mood. These sliders are either annotated with a color gradient and text or emojis to make the UI visually more appealing. After completing a task, the users are redirected to the dashboard screen.

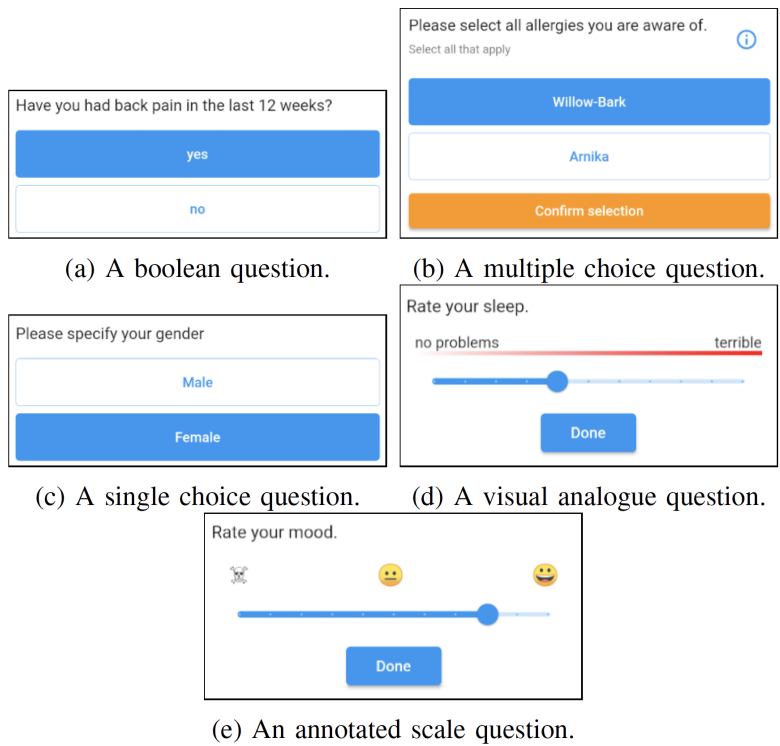

(a) A boolean question.  (b) A multiple choice question.

(c) A single choice question.  (d) A visual analogue question.

(e) An annotated scale question.

## Contact Screen

This screen is a hub for general information concerning the app. It references a Frequently Asked Questions (FAQ) screen with answers to basic user questions, a screen with contact information and the about screen.

## Report History Screen

The report history screen displays a list of studies in which the users have participated in the past. Selecting a study name redirects the users to the corresponding report screen.

## Report Overview Screen

We assume that knowledge about the previous course of complaints under certain interventions could influence the evaluation of complaints in the further course. Therefore, users should only have access to the results after a certain study period. Users should be able to access the results as soon as they reach the minimum study duration. However, it should be explained to users in advance that more reliable results can be expected the longer they remain participants in the study. We defined countable observations, which include only those that were completed on countable days and on which all adjacent interventions were completed.

Besides the performance report, the report overview includes the report sections defined by the researchers. One of the report types uses a linear regression model to determine which intervention improves the outcome. To achieve this, the desired outcome is used as the dependent variable. The independent variables are two dummy variables indicating whether intervention A or intervention B were active for the sample. The time since the start of the study is included to correct for a linear trend. Values predicted by this model are displayed with 95% confidence intervals in a bar chart. A hypothesis test is performed whether the regression coefficient of the intervention is zero or not using the large-sample Wald test statistic and a significance level of 0.05. A textual description of the comparison is provided in the report as well.

## Performance Screen

Users might wonder why the progress bars are not filling-up, e.g., if they just complete the observations but skip interventions (which would not be countable observations). Therefore, we built the performance details screen that shows for every intervention and observation how many were completed with the aim of reaching the suggested study length.

## Settings Screen

This screen allows the user to configure basic app functionality. It displays the active study name and enables them to change the language, opt-out/abort the current study and delete the user data on the device. The currently supported languages are German and English and concern only the UI elements of the app itself. Adding new languages to the app is very simple because the text strings are collected in a single file per language. Opting out of a study deletes the user's progress of the current study and takes them back to the study selection. Deleting the user data removes the user's anonymous account from the device and server and redirects them to the welcome screen. Data for studies that were finished by the users is kept on the server and not deleted.

# Supplementary Figures

*Supplementary Figure 1: Definition of interventions in the designer and selection of interventions in the app.*

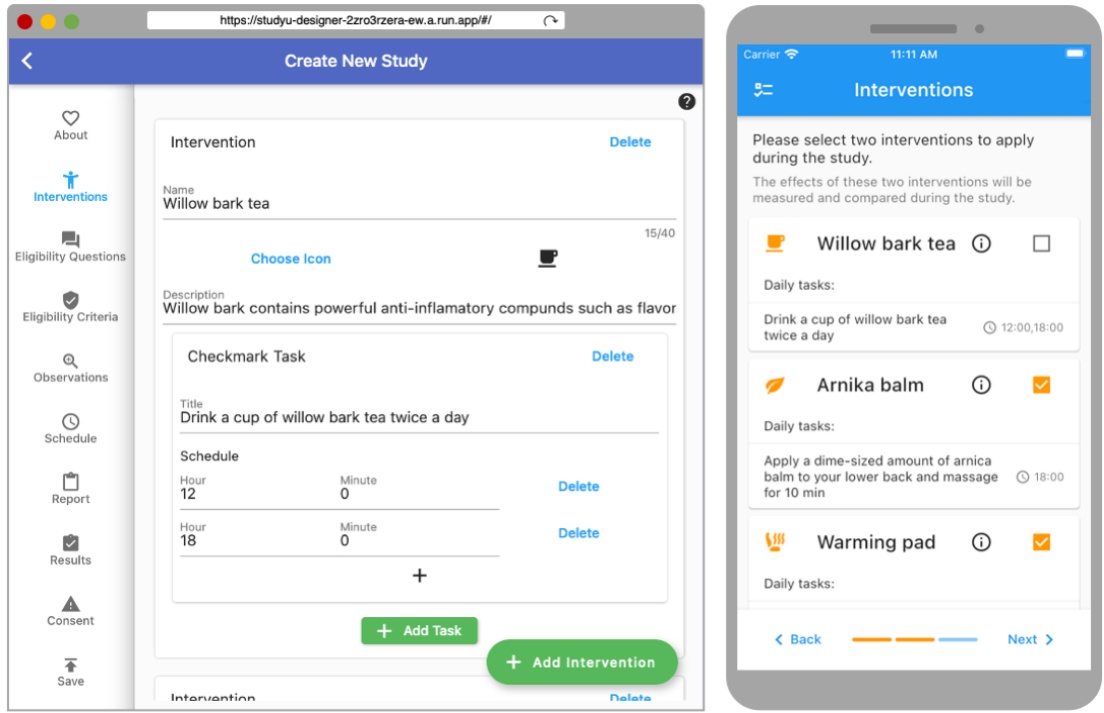

*Supplementary Figure 2: Eligibility questions and criteria in the designer and eligibility screen in the app.*

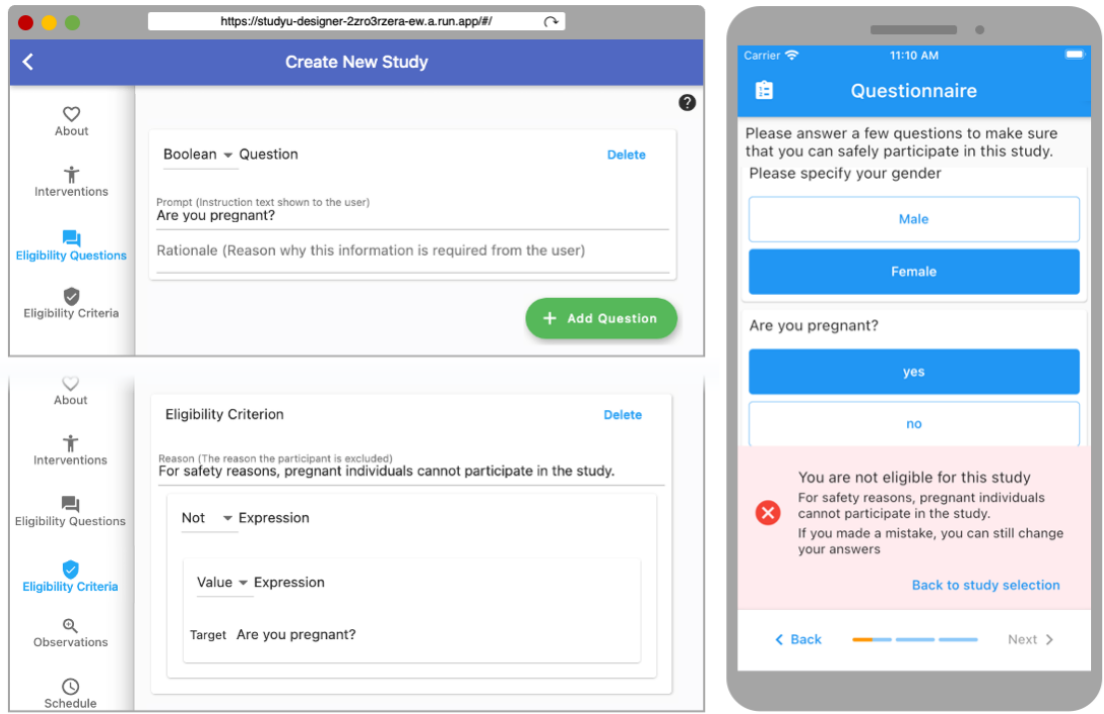

*Supplementary Figure 3: Definition of observations in the designer and the screens for the daily overview and outcome reporting in the app. In the example, the participant has to rate the individual pain level and mood on a Likert scale as well as medication intake.*

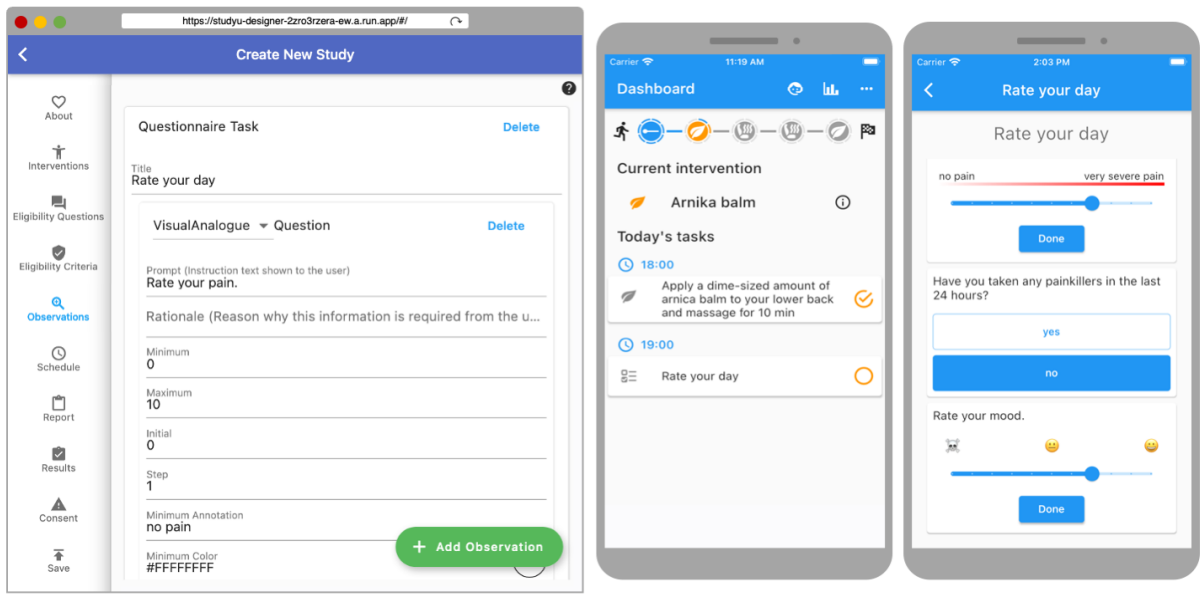

*Supplementary Figure 4: Consent definition in the designer and consent screen in the app. When selecting one consent item, details are displayed.*

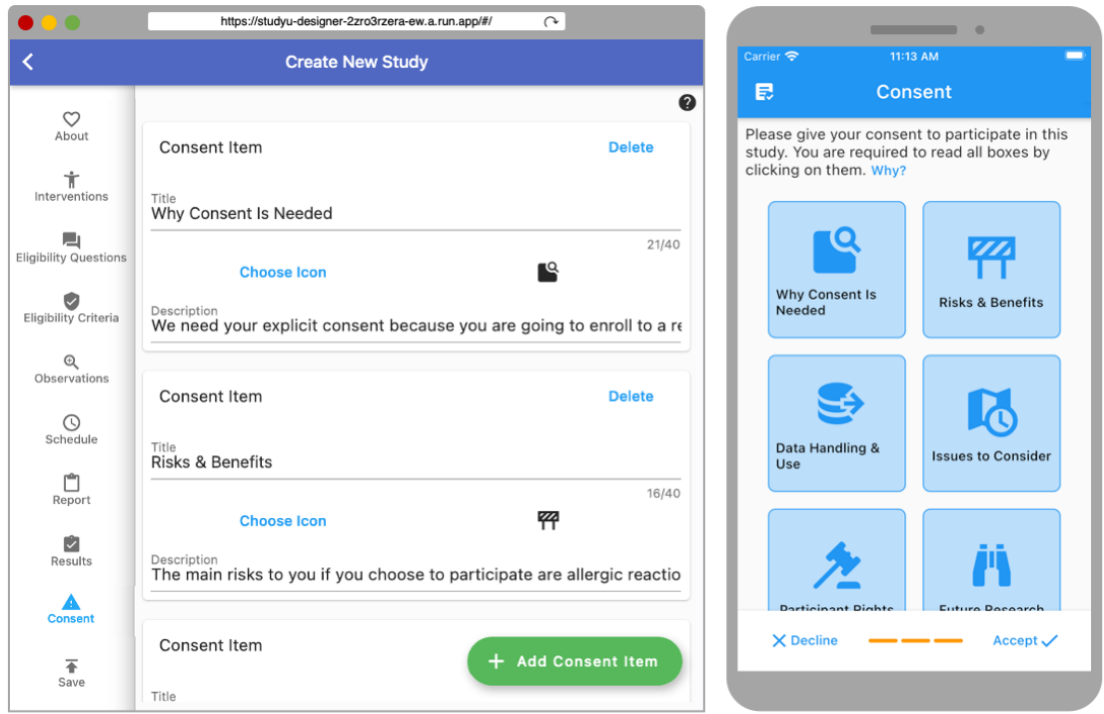

*Supplementary Figure 5: Overview of the full StudyU study model. The notation is based on the Unified Modeling Language (UML) class diagram notation: it defines properties of single classes in rectangles and associations between multiple classes as connections. The associations shown in this diagram with a filled diamond at one end mean that one class, e.g., 'Study', is composed of another class, in this case StudyDetails. Numbers shown at associations indicate how many instances of one class take part in this association, e.g., n 'Observation' objects can be associated with one 'StudyDetails' object.*

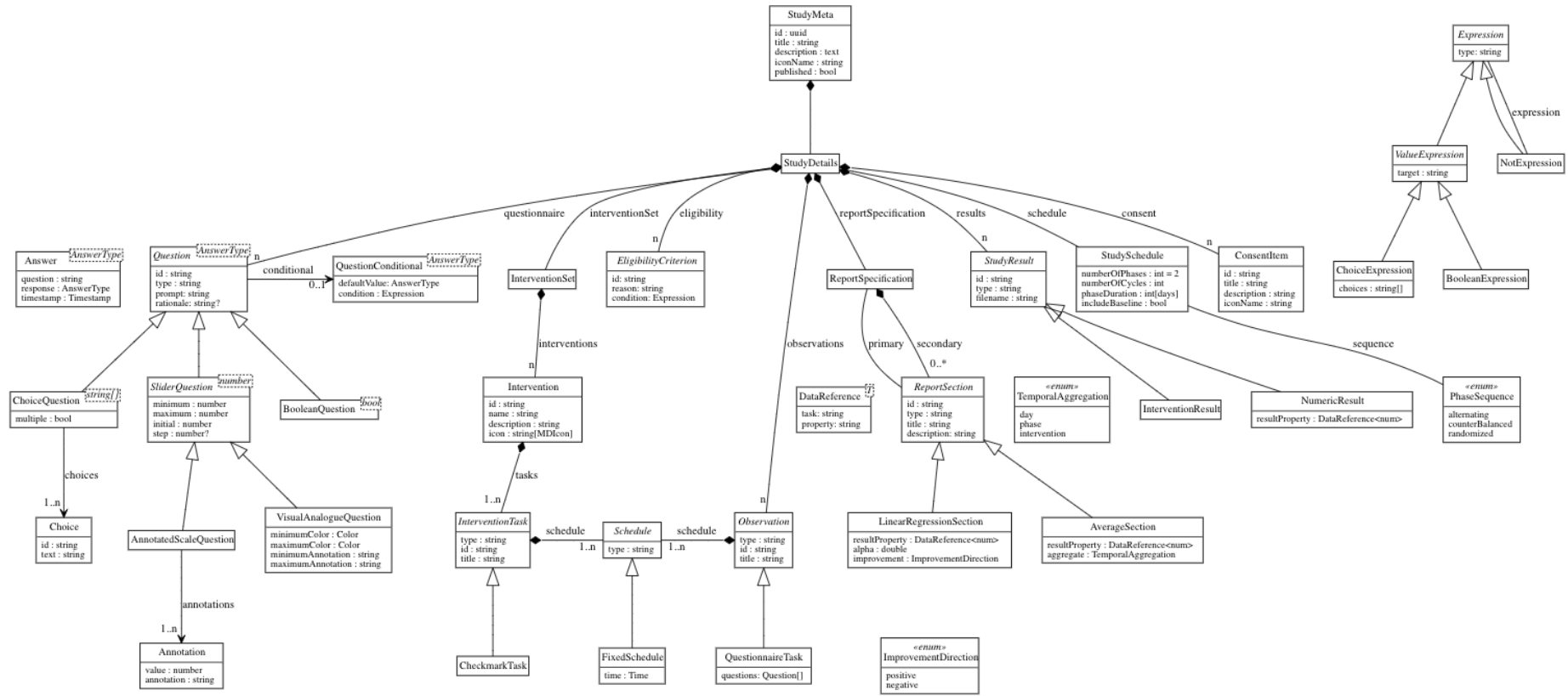